\documentclass[12pt]{article}
\usepackage{graphicx,epsfig,epsf,amssymb}
\usepackage{times}
\def \<{\langle}
\def \>{\rangle}

\title {Constructing a Robust Universe\\ with Attraction-Repulsion Coupling\\ and Energy Conservation}
\author
{Ti-Pei Li$^{1,2}$}
\date{}

\begin{document}
\baselineskip 24pt
\maketitle

{\footnotesize
\begin{enumerate}
\item{Department of Physics \& Center for Astrophysics, Tsinghua University, Beijing, China}
\item{Key Laboratory of Particle Astrophysics, Institute of High Energy Physics, Chinese Academy of Sciences, Beijing, China}
\end{enumerate}
}

\abstract{The discovery of accelerated cosmic expansion  implies that, in addition to the attractive gravity of matter, there exists in our universe some other form of energy (cosmological constant) producing a repulsive force. However, the cosmological constant expected by the quantum field theory is 120 orders of magnitude larger than what allowed by cosmological observations, which is called the cosmological constant problem and remains one of the most significant unsolved problems in fundamental physics. Here we show that the huge discrepancy can be resolved by assuming that our universe is an attraction-repulsion coupled system with energy conservation, and that the pre-inflation vacuum is in equilibrium between attraction and repulsion (in flat spacetime, not de Sitter or anti de Sitter). The attraction-repulsion coupling picture can also easily explain why both kinds of energy in our universe have similar magnitude today, and avoid singularity problems in general relativity and cosmology.  
}

\section{Introduction}
To get a stationary solution in cosmological situation from the field equation for general relativity (GR), Einstein~\cite{ein17} added an extra term proportional to the metric tensor $g_{\mu\nu}$ with a cosmological constant $\Lambda$, the Einstein field equation becomes
\begin{equation}\label{eq:fe}
G_{\mu\nu}+\Lambda g_{\mu\nu}=8\pi GT_{\mu\nu}\,,
\end{equation}
where, on the left-hand side (the geometry side), $G_{\mu\nu}$ is the Einstein tensor and, on the right-hand side (the physics side), $T_{\mu\nu}$ is the energy-momentum tensor and $G$ is the Newton's gravitational constant. Here we use the metric convention (-,\,+,\,+,\,+) and natural units $c=\hbar=1$. When the expansion of the universe was established, Einstein set the constant $\Lambda=0$. But the late discovery of an accelerating expansion rate by supernovae observations~\cite{rie98,per99} requires a positive non-zero $\Lambda$ to represent the cosmic repulsion responsible for the acceleration: the cosmological constant is interpreted as vacuum energy with a density 
\[ \rho_{_\Lambda}=\Lambda/8\pi G \]
 and a negative pressure 
\[ p_{_\Lambda}=-\rho_{_\Lambda}\,. \]
However, the density of vacuum energy estimated from the Planck energy is $\sim 10^{120}$ times larger than observational limits of dark energy, which is the famous cosmological constant problem in fundamental physics\cite{wei89}. 

The accelerated cosmic expansion implies that, in addition to gravitational attraction, cosmic repulsion also plays an essential role in the universe evolution: the cosmic attraction and repulsion both are intrinsic properties of spacetime, both have to be introduced to describe different eras of cosmic evolution. In doing so, we can construct an universe evolution picture with attraction-repulsion coupling to satisfy total energy conservation and get rid of the cosmological constant problem and other long-time puzzles e.g. the black hole singularity in GR and Big Bang singularity in cosmology.\\

\section{Attraction-Repulsion Coupling}
\subsection{Before inflation}
Firstly, we consider the universe before the creation of matter. For the original vacuum before inflation, $T_{\mu\nu}=0$, the Einstein field equation~(\ref{eq:fe}) becomes
\begin{equation}\label{eq:vfe}
G_{\mu\nu}+\Lambda g_{\mu\nu}=0\,.
\end{equation}
However, the field equation~(\ref{eq:vfe}) of de Sitter universe with a positive cosmological constant term responsible only for the cosmic repulsion is obviously not a complete description for the attraction-repulsion coupled universe. For the pre-inflation vacuum, it is needed to use two constants, $\Lambda_r$ and $\Lambda_a$, responsible for repulsion and attraction respectively, and the vacuum equation should be
\begin{equation}\label{eq:vfe1}
G_{\mu\nu}+(\Lambda_r-\Lambda_a)\,g_{\mu\nu}=0\,.
\end{equation}

In the vacuum universe before inflation, the cosmic repulsion and attraction compensate each other. In comparing with attraction/repulsion domination, assuming a balanced flat spacetime with the repulsive constant being equal to the attractive one,  $\Lambda_r=\Lambda_a$, is a natural choice for the very early universe before the creation of matter, while the composite cosmological constant $\Lambda=\Lambda_r-\Lambda_a$ in Eq.~(\ref{eq:vfe1}) is zero, which is just what Einstein favored. It has to be pointed out that the vacuum field equation~(\ref{eq:vfe}) with Einstein's setting ($\Lambda=0$) is not identical to Eq.~(\ref{eq:vfe1}) with $\Lambda_r-\Lambda_a=0$ (or $\Lambda_r/\Lambda_a=1$): the later still contains energies from attractive and repulsive fields, although the two fields are compensated in interaction with each other and their energies are bounded. 

The attraction-repulsion coupling vacuum described by Eq.~(\ref{eq:vfe1}) with a zero composite cosmological constant $\Lambda$, is neither de Sitter nor anti de Sitter, but a static flat spacetime. The two constants in Eq.~(\ref{eq:vfe1}), $\Lambda_r$ and $\Lambda_a$, are proportional to the repulsive pressure $-p_{_r}$ and the attractive pressure $p_{_a}$, respectively. A spacetime with the two mutual coupled compensation fields and with energy conservation can keep static or quasi-static because the energy conversion between the two fields could help to damp expansion or contraction. Inhomogeneities induced by local fluctuations can be smoothed by energy transmission in the vacuum. Therefore, a spring-like mechanism for a repulsion-attraction coupled system could help avoid big collapse or expansion and keep the vacuum spacetime static and flat in overall perspective. 

\subsection{Inflation}
If at a moment $t=t_i$ a region of radius $R_i$ has $\Lambda\gg 0$ caused by, e.g., enough large number subregions with $\Lambda_r\gg \Lambda_a$ randomly by fluctuation, an inflation can be triggered like a continuous phase transition between two states. The large positive net cosmological constant breaks the coupling of repulsion and attraction fields and drives the de Sitter-type vacuum to start accelerated exponential expansion. At the end of inflation, matter is generated from the cosmic attraction field, accordingly the constant $\Lambda_a$ term in the left-hand side of the vacuum field equation Eq.~(\ref{eq:vfe1}) is transferred to the term of energy-momentum tensor $T_{\mu\nu}$ at the right-hand side of the Einstein's field equation. To keep the total energy conserved, at inflation end, $t=t_m$, the parameter in the dark energy term should be 
\begin{equation}\label{eq:lambda}
\lambda(t_m)=\Lambda_r/[a(t_m)]^3
\end{equation}
with $a(t)=R(t)/R(t_i)$ being the time-dependent dimensionless scale factor normalized to the inflation starting epoch $t_i$ by $a(t_i)=1$.  Therefore, the inflation in our scheme is a transient process between the original vacuum and present universe, which is driven by the cosmic repulsive field to convert the cosmic attractive and repulsive energies bounded in the vacuum into the energies of matter and dark energy appeared in the post-inflation universe, or, in other words, to transfer the two cosmological constant terms in the vacuum field equation~(\ref{eq:vfe1}) from the geometry side to the physics side to be two terms with the energy-momentum tensor and dark energy parameter. When the potential vacuum energy for repulsion is completely converted, the inflation has to come to an end.

The inflation as a transition process is hard to be described by a GR equation. We can use the following simplified equation to intuitively but roughly illustrate it: 
\[ G_{\mu\nu}+[(\Lambda_r-\Delta\Lambda_r)/a^3]g_{\mu\nu}-(\Lambda_a/a^2)g_{\mu\nu}=-\Delta\Lambda_r/a^3\,. \] 
The inflation stops at $t_m$ when the potential vacuum energy for repulsion is completely converted, $\Delta\Lambda_r=\Lambda_r$,  and then the term with $\Lambda_a$ of the left-hand side is converted into $8\pi G T_{\mu\nu}$ of the right-hand side, the field equation becomes Eq.~(\ref{eq:fe1}).

\subsection{After inflation}
The energy-momentum tensor of a homogeneous universe with both matter and dark energy is
\[
\tilde{T}_{\mu\nu}=(\tilde{\rho}+\tilde{p})u_\mu u_\nu+\tilde{p}g_{\mu\nu}\,,
\]
where $u^\mu=(1,0,0,0)$, $\tilde{\rho}$ and $\tilde{p}$ are respectively the total energy density and the pressure of the cosmic medium: $\tilde{\rho}=\rho_m+\rho_{_\lambda}$ and $\tilde{p}=p_m+p_{_\lambda}$ with the suffixes $m$ and $\lambda$ indicating "matter" and "dark energy", respectively. For the dark energy,  
$p_{_\lambda}=-\rho_{_\lambda}$, and the field equation can be written as
\begin{eqnarray}\label{eq:fe1}
G_{\mu\nu} & = & 8\pi G\tilde{T}_{\mu\nu} \nonumber \\
          & = & 8\pi GT_{\mu\nu}-\lambda g_{\mu\nu}\,,
\end{eqnarray}
where the matter's energy-momentum tensor 
\[ T_{\mu\nu}=(\rho_{_m}+p_{_m})u_\mu u_\nu+p_{_m}g_{\mu\nu} \]
 and 
\[ \lambda=8\pi G \rho_{_\lambda}=-8\pi G p_{_\lambda}\,. \]
 For the cosmic medium of radius $R$ in the co-moving frame, from the local energy-momentum conservation $\nabla^{^\mu} \tilde{T}_{\mu\nu}=0$, we get
\begin{equation}\label{eq:dR}
\frac{d}{dR}\{[\rho_{_m}(t)+\rho_{_\lambda}(t)] R^3(t)\}=-3[p_{_m}(t)+p_{_\lambda}(t)]R^2\,.
\end{equation}
For an attraction-repulsion balanced universe, $p_{_m}(t)+p{_\lambda}(t)=0$, from Eq.~(\ref{eq:dR}) we can derive the total energy conservation law
\begin{equation}\label{eq:ec}
[\rho_{_m}(t)+\rho_{_\lambda}(t)] R^3(t)=\mbox{constant}\,.
\end{equation}

During the universe's expansion, the energy density of matter and that of dark energy are both decreasing. To keep the total energy conserved, the two kinds of energies should be converted into each other: in phases of dark energy-driven acceleration, energy should be converted from dark energy to matter, and inversely from matter to dark energy in phases of deceleration driven by gravity. During matter creation in the expanded spacetime at the end of accelerating inflation, matter's rest mass can be generated from the cosmic attractive field by, e.g., a Higgs-like mechanism, but the kinetic energy of rapidly expanded matter has to much expend dark energy, the universe may become matter dominated at a time $t_d$ and enters into a decelerating phase, which is a transition period between the de Sitter-type inflation at incredibly accelerated speeds and the present slightly accelerated universe. During the deceleration phase, the fractional dark energy is increasing with expansion as kinetic energy of matter being inversely converted back to it, until the universe becomes dark energy dominated at a time $t_a$ and begins accelerating again. In the future, the universe could enter into the next deceleration phase at some time $t_{d'}$,  and through acceleration-deceleration cycles should finally reach a new equilibrium state,  a gravity and dark energy balanced state at a radius $R_f$. In this picture, the present universe is just a temporary transition from the no-beginning static vacuum universe to an endless static or stationary universe in the long-term future, as shown schematically in Figure 1.

\begin{figure}[t]
\label{f1}
   \begin{center}
\includegraphics[height=8cm, width=6cm, angle=270]{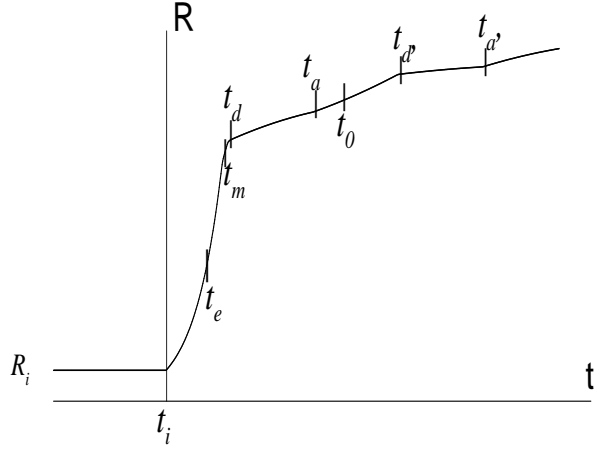}
   \caption{\footnotesize A schematic depiction for expansion history of an attraction-repulsion coupled and energy conserved universe. Marked times: $t_i$ -- inflation beginning, $t_e$ -- generation of electromagnetic, $t_m$ -- matter creation, $t_d$ -- starting the deceleration phase, $t_a$ -- starting the acceleration phase, $t_0$ -- the present time, $t_{d'}$ -- deceleration again, $t_{a'}$ -- acceleration again. For clarity proposes the figure is drawn not to scale.
}
   \end{center}
\end{figure}

\section{Troubles in GR and Comology}  
That an integral energy-momentum conservation law cannot be derived from the local conservation equation when gravity is present (spacetime curved) has been an awkward problem and evoked strong objections to the GR theory~\cite{tra62,pais82}. To simultaneously take into account both the attractive field and the repulsive field can help to ease this problem. The two equilibrium eras in our scheme, the pre-inflation universe with $\Lambda_r-\Lambda_a=0$ in Eq.~(\ref{eq:vfe1})  and the future universe with $p_{_m}+p_{_\lambda}=0$ in Eq.~(\ref{eq:dR}), are both equilibrium of attraction and repulsion with a flat Minkowski space and both follow integral conservation law for energy and momentum. Even for the transitory expansion era, the presence of the cosmic repulsive field can compensate gravity to make the universe close to being flat, much alleviating the problem of integral conservation. In fact, similarly to a spring coupled system, it is possible to construct a relaxation process to a gravitation-repulsion coupled universe with energy conversion and conservation to reach a final equilibrium state. There is no necessary reason to break the fundamental physical law. Therefore, the total energy conservation described by Eq.~(\ref{eq:ec}) could be taken as a prerequisite constraint in studying the dynamics of our universe.

There exist two kinds of vacuum in our picture: the original vacuum described by Eq.~(\ref{eq:vfe1}) which is a false vacuum with both attractive and repulsive cosmic energies represented by two cosmological constants $\Lambda_a$ and $\Lambda_r$ respectively, and the post-inflation vacuum with only dark energy $\lambda$ appearing in the right side of Eq.~(\ref{eq:fe1}). After inflation, the vacuum has exhausted its attractive energy and its repulsive energy density is much weakened and thus cannot produce inflation with matter creation anymore. The discrepancy of $\sim 120$ orders of magnitude between the theoretical expectation of the cosmological constant $\Lambda_r$ in Eq.~(\ref{eq:vfe1}) and the observational value of dark energy $\lambda$ in Eq.~(\ref{eq:fe1}) for the present time can be easily explained: it is produced mainly by the extremely huge inflation. From Eq.~(\ref{eq:lambda}), if the universe expanded by a factor of $a(t_m)\simeq 10^{40}$ during inflation, the dark energy density at the end of inflation should be only $\simeq 1/10^{120}$ of the vacuum energy density for the pre-inflation repulsive field. 

 Singularity problems in GR and cosmology can be also solved by our picture. With gravitation only, Einstein's GR laws predict singularities inside black holes, this is the reason for Einstein to strongly reject black holes. However, for a black hole from the GR field Eq.~(\ref{eq:fe1}) with space scale $R$, the dark energy density, then the repulsive pressure, is increasing with spacetime collapse as $R^{-3}$, whereas the gravitational strength is increasing slowly as $R^{-2}$, the collapse should stop when the repulsive pressure comes stronger than the gravitational pressure. Moreover, after formation of a black hole, the gravitational binding energy released during the continuous collapse might be converted into the coupled dark energy, which can brake the collapse earlier. Therefore, in an attraction-repulsion coupled universe a black hole ends not as a singularity but a balanced state between attraction and repulsion or oscillation around it. The Big Bang singularity also does not exist in our picture, because the observed universe is generated not from a singular point, but from a finite region of size $R_i$ in the original vacuum. No need for a Big Bang at all. 

The thermal history of the very early universe in our scenario is completely different with the Hot Big Bang picture. In the Hot Big Bang scenario, after an extremely hot Planck era the temperature drops down with universe expanding to below $\sim 10^{12}$\,K ($\sim 10^{15}$\,GeV) and the GUT (Grand Unified Theories) phase transition occur. The gravitation separates from the other forces of nature, then the strong, weak and electromagnetic forces further separate when the universe crosses a series of symmetry-breaking (phase-transition) temperatures in succession. However, the Planck and GUT eras do not appear in our picture, the universe is originated from a cold vacuum spacetime and heated later during inflation. The recent result from re-analyzing the Wilkinson Microwave Anisotropy Probe (WMAP) data with removing the scan-induced artificial anisotropy shows the quadrupole of the cosmic microwave background (CMB) anisotropy being almost zero~\cite{lit09, liu10, liu11a, liu11b}, i.e. almost no CMB fluctuation being detected on the largest angular scale. Such a remarkable observation fact indeed suggests a very homogeneous and cold state for the very early universe. Consequently, the history of field separation and particle creation in the very early universe may be much different from what expected from the Hot Big Bang.  The electromagnetic field can separate from the cosmic attractive field earlier than the gravity, the later should be generated along with matter via a Higgs-like mechanism at the end period of inflation. Such an order for separation of the two long-distance forces may help us understand the origin of so called cosmic large numbers, a century long subject in physics revealed and explored by Weyl~\cite{wey18}, Eddington~\cite{edd31}, Dirac~\cite{dir37,dir38} and constant successors~\cite{pen04}. For example, if the strength of gravitational/electromagnetic force reflects the strength of the original attractive field at the epoch it separated out and if the strength of the attractive field strength changes with scale by $a^{-2}$, the Newton's gravitational constant $G$ should depend on the cosmic curvature at time $t_m$, i.e. on $[a(t_m)]^{-2}$,  and the one of the cosmic large numbers, the intensity ratio of electromagnetic and gravitational interaction $e^2/Gm_e^2\sim 10^{42}$, can be interpreted by the electromagnetic field separation time $t_e$ being earlier than the gravitational field with the ratio of two scale factors $a(t_m)/a(t_e)\sim 10^{21}$. 

\section{Discussion}
The instability in GR and various universe models is an essential source of troubles in fundamental physics. The instability mainly comes from taking into account only attraction or only repulsion in the field equation, e.g. only gravity in $G_{\mu\nu}=8\pi GT_{\mu\nu}$ for Einstein's universe, only repulsion in $G_{\mu\nu}+\Lambda g_{\mu\nu}=0$ for de Sitter universe, only attraction in $G_{\mu\nu}-\Lambda g_{\mu\nu}=0$ for anti-de Sitter universe. The field equation $G_{\mu\nu}=8\pi GT_{\mu\nu}-\Lambda g_{\mu\nu}$ for the Lemaitre universe although includes both gravitation (the energy-momentum tensor $T_{\mu\nu}$) and repulsion (the cosmological constant $\Lambda$) but not considers their coupling, also exists the gravitational instability~\cite{nar69}. To construct an attraction-repulsion coupled universe we use two field equations:
\begin{eqnarray*} 
\lefteqn{\hspace{-70mm} G_{\mu\nu}-\Lambda_a g_{\mu\nu}+\Lambda_r g_{\mu\nu}=0\hspace{2mm}(\Lambda_a=\Lambda_r) \hspace{5mm}\mbox{for pre-inflation,}}\\
\hspace{-40mm} G_{\mu\nu}=8\pi GT_{\mu\nu}-\lambda g_{\mu\nu} \hspace{5mm} \mbox{for post-inflation.}
 \end{eqnarray*}
Here the flat pre-inflation false vacuum is combined by the positive curved de Sitter spacetime and negative curved anti-de Sitter spacetime, i.e., constituted with repulsive and attractive fields containing their energies and balancing in pressure. The energies in the pre-inflation vacuum are the source of the matter/radiation energy and dark energy in the post-inflation universe. The time dependence of the dark energy parameter $\lambda$ in the field equation for post-inflation universe is determined by the conversion of the two classes of energy and the total energy conservation.

In our scenario, the present universe begin from a finite region of $R_i$ in an infinite vacuum field, which is constituted by two balanced fundamental cosmic fields: the cosmic attractive field and the cosmic repulsive field. Fluctuation and decoupling of the two fields in the primordial region of our universe causes its spacetime inflation starting at $t_i$ (the original cosmic vacuum has no beginning, we can arbitrarily set a zero point for time).  The universe is mostly static or quasi-static during its infinite life: in the no beginning flat vacuum with balanced cosmic attraction-repulsion fields or in the future endless gravity-dark energy balanced state, where the present universe is just a brief transition between the two equilibrium states. In this picture, our universe is just a specific physical object, an finite attraction-repulsion coupled mechanical system with limited energy, no singularity and obeys the energy conservation law. After understanding the properties of the two cosmic fields and their interaction and conversion, based on cosmological observations we can evaluate the parameters of our universe: the cosmological constant $\Lambda_a$ for attraction, $\Lambda_r$ for repulsion, the initial scale $R_i$, and the final scale $R_f$, etc.  Although in our scenario eternal inflation cannot occur from a  post-inflation vacuum, but the existence of parallel universes is its natural deduction: besides our universe, different universes can be parallel-generated from the original vacuum at different locations and different times with different initial parameters.  From studying physics of the fundamental cosmic fields, it is possible to deduce profiles of other universes statistically.

\end{document}